\documentclass[11pt,iop]{emulateapj}

\usepackage{apjfonts}

\shorttitle{{\rm GeV} emission from starburst galaxy NGC~2146}
\shortauthors{Tang et al.}

\begin{document}

\title{Discovery of {\rm GeV} emission from the direction of the
luminous infrared  galaxy NGC~2146}

\author{Qing-Wen Tang\altaffilmark{1,2}, Xiang-Yu Wang\altaffilmark{1,2} and Pak-Hin Thomas Tam\altaffilmark{3}}
\affil{$^1$ School of Astronomy and Space Science, Nanjing University, Nanjing, 210093, China; \\ xywang@nju.edu.cn  \\
$^2$ Key laboratory of Modern Astronomy and Astrophysics (Nanjing University), Ministry of Education, Nanjing 210093, China \\
$^3$ Institute of Astronomy and Department of Physics, National Tsing Hua University, Hsinchu 30013, Taiwan  \\ phtam@phys.nthu.edu.tw\\
}

\begin{abstract}
Recent detection of high-energy gamma-ray emission from starburst
galaxies M82 and NGC 253 suggests that starburst galaxies are huge
reservoirs of cosmic rays and these cosmic rays convert a
significant fraction of their energy into  gamma-rays by colliding
with the dense interstellar medium. In this paper, we report the
search for high-energy gamma-ray emission from several nearby
star-forming and starburst galaxies using the 68 month data
obtained with the Fermi Large Area Telescope.  We found a
$\sim5.5\sigma$ detection of  gamma-ray emission above 200{\rm
MeV} from a source spatially coincident with the location of the
luminous infrared galaxy NGC~2146. Taking into account also the
temporal and spectral properties of the gamma-ray emission, we
suggest that the gamma-ray source is likely to be the counterpart
of NGC~2146. The gamma-ray luminosity suggests that cosmic rays in
NGC~2146 convert most of their energy into secondary pions, so
NGC~2146 is a "proton calorimeter". It is also found that NGC~2146
obeys the quasi-linear scaling relation between the gamma-ray
luminosity and total infrared luminosity for star-forming
galaxies, strengthening the connection between massive star
formation and gamma-ray emission of star-forming galaxies.
Possible TeV emission from NGC~2146 is predicted and the
implications for high-energy neutrino emission from starburst
galaxies are discussed.
\end{abstract}


\section{Introduction}

It is generally believed that Galactic cosmic rays (CR) are
accelerated by supernova remnant (SNRs)
shocks~\citep{Ginzburg1964}. CR protons interact with the
interstellar gas and produce neutral pions (schematically written
as $p+p\rightarrow \pi^0+{\rm other \, products}$), which in turn
decay into gamma-rays ($\pi^0\rightarrow\gamma+\gamma$). The high
SN rate in starburst galaxies implies high CR emissivities, so
they are predicted to be bright gamma-ray sources (e.g.
\citet{Torres2005,Rephaeli2010,Lacki2010,Ackermann2012}). The
gamma-ray luminosity of starbursts depends not only on the CR
intensity, but also on the conversion efficiency of CR proton
energy into pionic gamma-rays. This efficiency in turn depends on
the ratio of the timescale of pion production to the escape time
of protons. Protons escape by advection in galactic winds or by
diffusion.  A galaxy becomes a "proton calorimeter" when the pion
production time is shorter than the escape time , i.e. CR protons
lose almost all of their energy to pionic collisions before
escaping \citep{Thompson2007}. The detection of high-energy
gamma-ray emission from starburst galaxies M82 and NGC 253 suggest
that CR protons convert a significant fraction of their energy
into secondary pions \citep{Abdo2010a}, so they are close to be
"proton calorimeter" ~\citep{Lacki2011}.

\citet{Ackermann2012} (hereafter ACK12)  examined a sample of 64
dwarf, spiral, and luminous and ultra-luminous infrared galaxies
using 3 years of data collected by the Large Area Telescope (LAT)
on the Fermi Gamma-ray Space Telescope. With a larger sample of
significantly detected sources and flux upper limits for the
remaining galaxies,  they found that the gamma-ray luminosity and
total infrared luminosity of star-forming galaxies are correlated.
The obtained scaling relations  allow straightforward predictions
for the next star-forming galaxies that could be detected by
Fermi/LAT. Seven galaxies are selected by ACK12 as the best
candidates with the highest probabilities for the next Fermi/LAT
detection beyond the Local Group: M83, NGC 3690, NGC 2146, Arp
220, NGC 1365, M51, and M101.

We search for high-energy gamma-ray emission from the above 7
galaxies using the 68 month Fermi/LAT data. We found significant
detection of gamma-ray emission from NGC~2146, the nearest
luminous infrared galaxies (LIRGs). LIRGs are extreme starburst
galaxies with total infrared luminosity in 8-1000 $\mu {\rm m}$
 $\ga10^{11}L_\odot$~\citep{Sanders1996}.
They have very high star formation rate (SFR) in concentrated
regions. The high infrared luminosities are due to large amounts
of dust, which absorb ultra-violet (UV) photons and re-radiate
them in the infrared (IR).   NGC~2146 has $L_{\rm 8-1000 \mu {\rm
m}}\simeq 10^{11}L_\odot$ at distance of $D=15.2{\rm
Mpc}$~\citep{Gao2004}. For the other 6 galaxies, upper limits on
the gamma-ray flux are obtained.

\section{LAT data analysis and results}

\subsection{Data reduction and analysis}
\label{method} Our sample comprises 7 star-forming galaxies which
are selected by ACK12 as the most promising sources for Fermi/LAT
detection. Here we present the data derived from the Fermi/LAT
observations from 2008 August 4(MJD 54, 682) to 2014 April 4(MJD
56,752). Photons with energy between 200{\rm MeV} and 100 {\rm
GeV} are selected to be analyzed when using the Fermi Science
Tools package, version v9r32p5. Instrument response
functions(IRFS) of P7REP$\_$SOURCE$\_$V15 is used for these events
of source class. Also, we excluded the events with zenith angles
$>$100$^\circ$ in order to avoid a significant contribution of
Earth-limb gamma-rays. All events in a region of interest (ROI) of
10$^\circ$ around the positions of galaxies have been used, and
all sources within 15$^\circ$ listed in the second Fermi
catalog~\citep{Nolan2012} (2FGL catalog) were modeled to produce
the spectra and calculate the test statistic (TS, see e.g.,
~\citet{Mattox1996}) of the source. The diffuse Galactic and
extragalactic background components were modeled with the files
gll$\_$iem$\_$v05.fits and iso$\_$source$\_$v05.txt, respectively.
All galaxies were modeled with a power law spectra
$N(E)=N_0(E/E_0)^{-\Gamma}$, where $\Gamma$ is the photon index.
To assess the detection significance of gamma-ray emission for
galaxies, unbinned likelihood analysis are performed with the
model described above. If the corresponding TS value is above 25,
we derived the best-fit position using the tool {\it gtfindsrc}.
For the sources included in the 2FGL catalog, we fixed their
position to the values from the catalog.  Then we perform the
maximum likelihood analysis again to refine the spectral
parameters of the sources within the ROI and the diffuse sources
using the {\it gtlike}. Binned likelihood analysis was also
performed  to verify the results.

\subsection{NGC 2146}
\begin{figure}
    \epsscale{.8}
   \plotone{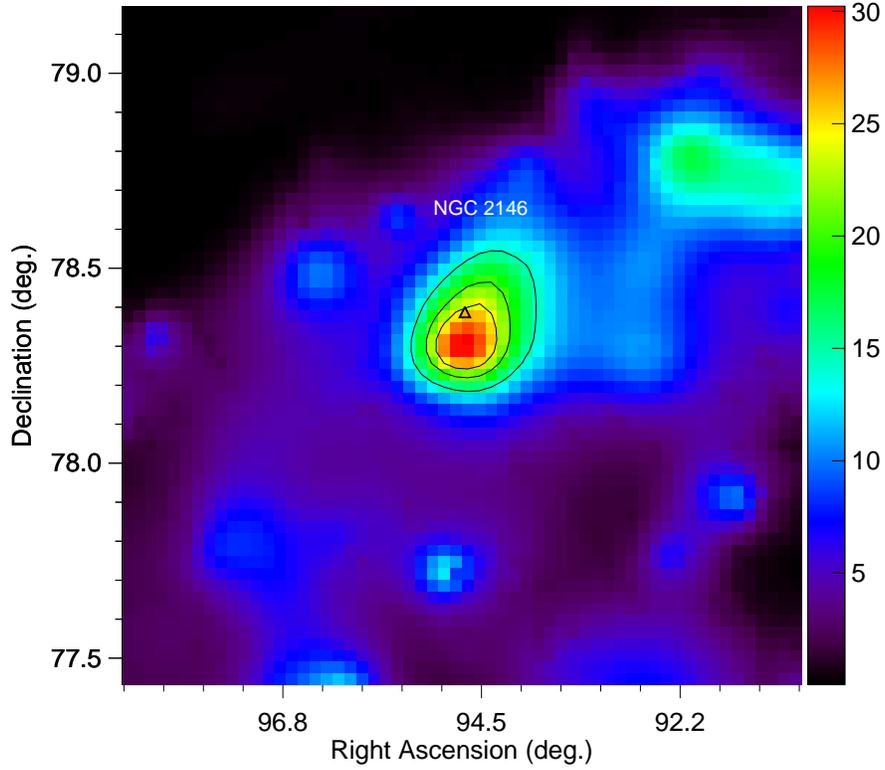}
   \caption{Test statistic map obtained from photons above 200 {\rm MeV} showing the celestial
   region (1.8$^\circ$ by 1.8$^\circ$, a pixel with 0.03$\times$0.03 squared-degrees) around NGC~2146.
   The black empty triangle indicates the optical position of NGC~2146; black lines show the 1$\sigma$, 2$\sigma$, and 3$\sigma$
   confidence level contours on the position of the observed gamma-ray excess. The color scale indicates the
   point-source TS value at each location on the sky. \label{tsmap}}
    \end{figure}

\begin{figure}
    \epsscale{1.0}
   \plottwo{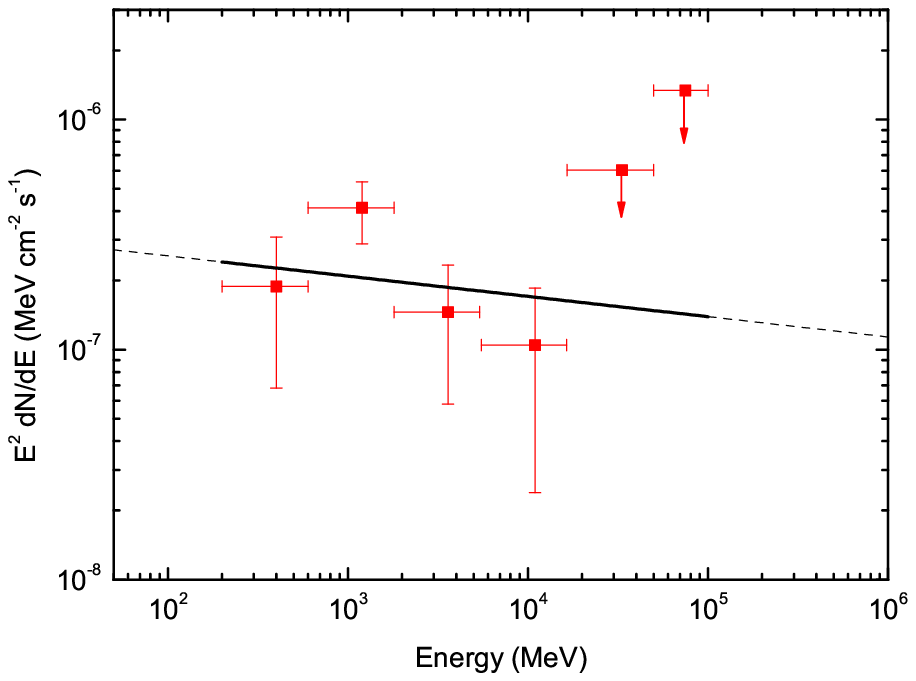}{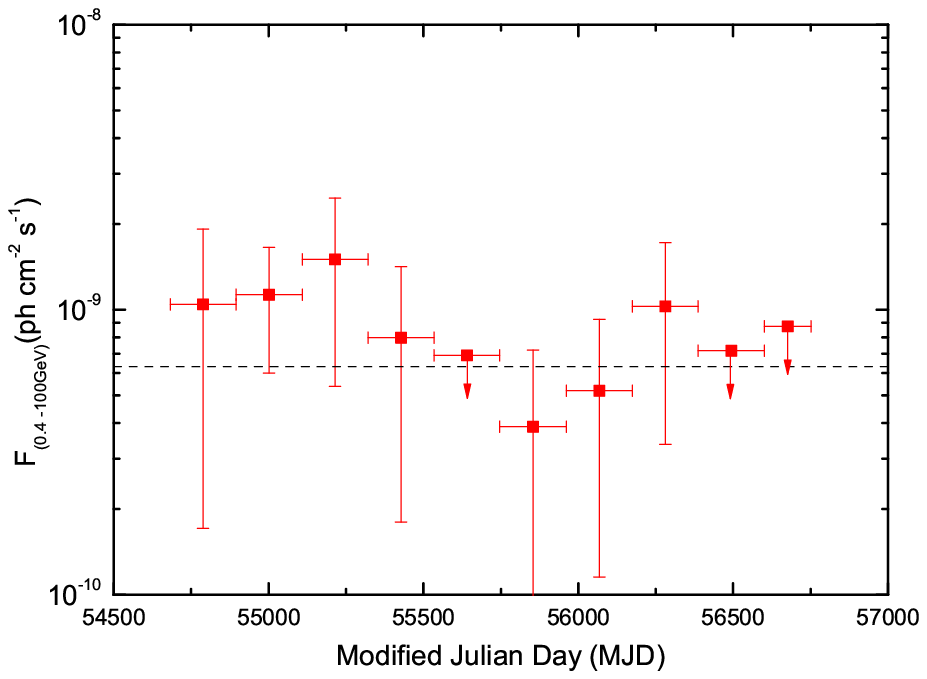}
   \caption{Left: Spectral energy distribution for NGC~2146 obtained in the analysis of the 68 months of data as described
   in the text.  The black solid line represents the best-fit power law in the range of 0.2--100~{\rm GeV}as shown in Table  \ref{position}.
   Right: Gamma-ray (>400 MeV)  light curve for NGC~2146  obtained in the analysis of the 68 months of data. The 68 month observation period was divided into 10 time
   intervals for analysis.
   The dashed black line shows the maximum likelihood flux level obtained for the full 68 months observation period. One-$\sigma$ errors are shown for energy bins with TS$>$1  and 95\% confidence-level upper limits are shown otherwise.}
   \label{splc}
   \end{figure}

A significant gamma-ray excess with TS of 30.8 (corresponding to a
detection significance $\approx$5.5$\sigma$) was found at the
best-fit gamma-ray position (R.A.,
Decl.)=(94.55$^\circ$,78.30$^\circ$). This position is consistent
with the core of NGC~2146, the nearest LIRG galaxy. Detection
significance map for NGC~2146 is shown in Fig. \ref{tsmap}. This
result unambiguously confirms an earlier, marginally significant
excess (TS$\sim$20) in the direction of NGC~2146 reported in
ACK12. The 0.2-100~{\rm GeV} spectrum of NGC~2146 is best
described by a power law with $\Gamma$=2.1$\pm$0.2 and a photon
flux of $F(0.2-100{\rm GeV})$=(1.1$\pm$0.5)$\times
10^{-9}$~cm$^{-2}$s$^{-1}$. The localization and spectral results
of the likelihood analysis are summarized in Table \ref{position}.
We also performed an additional maximum likelihood analysis by
dividing the energy range 0.2--100~{\rm GeV} into six energy bins,
and the spectrum is shown in Fig.~\ref{splc}.

We test alternative associations for the gamma-ray excess besides
NGC~2146 in the CRATES catalog~\citep{Healey2007} of flat-spectrum
radio sources (FSRQs), Veron-Cetty Catalog of Quasars \&
AGN~\citep{Veron2010} and the Candidate Gamma-Ray Blazar Survey
catalog, CGRaBS~\citep{Healey2008}. In the vicinity of NGC~2146,
the only possible gamma-ray emitting object is the FSRQ
CRATES~J061758$+$781552 (also named CGRaBS~J0617$+$7816 with
redshift of 1.43~\citep{Caccianiga2000}),  which is 0.04$^\circ$
away from best-fit gamma-ray position.

We searched for gamma-ray flux variability by creating a
7-month-per-bin light curve of the photon flux $>400$ {\rm MeV}
arriving from within a circular region of 10$^\circ$ in radius
centered on the   best-fit gamma-ray position,  which is shown in
Fig. \ref{splc}. Measuring the goodness-of-fit of the light curve
using the average flux over the years gives a reduced $\chi^2$ of
0.84 with 7 degrees of freedom, while upper limits were not
included. Lack of variability is consistent with a starburst
galaxy origin, while FSRQs with such a low variability in
gamma-rays is rare among LAT blazars. The photon index
$-2.1\pm0.2$ also agree with that of other gamma-ray detected
star-forming galaxies in ACK12, whereas the mean photon index of
LAT-detected FSRQs is $-2.46$~\citep{Abdo2010b}. Therefore, based
on the gamma-ray spectrum and light curve, we conclude that the
gamma-ray excess is dominantly contributed by NGC~2146, although a
small contribution by CRATES~J061758$+$781552 can not be formally
excluded.

\subsection{M51, NGC 1365, Arp 220 and M83}
Four more sources with significant gamma-ray excesses were found
in the vicinity of M51(TS=29.9), NGC 1365(TS=44.4), Arp 220(TS=52.2)
and M83(59.1); however, the best-fit positions are all far from the
core of the corresponding galaxies, i.e,  the angular separation
is larger than $r_\mathrm{95}$, c.f., Table~\ref{position}. We
therefore explored possible candidate sources for those gamma-ray
excess within $r_\mathrm{95}$.

M51.  The gamma-ray excess next to M51 likely comes from
SDSS~J13261$+$4754, with an angular separation of 0.13$^\circ$
between them, which is smaller than $r_\mathrm{95}$.

NGC~1365. The gamma-ray excess is likely associated with
PKS~0335$-$364, a strong radio quasar,  with an angular separation
of 0.07$^\circ$ between them.

Arp~220.  Both CRATES~J153246+234400 (a FSRQ) and SDSS~J15323+2333
(a Seyfert 2 galaxy at $z$=0.047) may contribute to the gamma-ray
excess since their angular separations are just 0.16$^\circ$ and
0.14$^\circ$, respectively. We further constructed an annual light
curve of the gamma-ray emission and found a significant
variability (a factor of 4 between the highest and lowest annual
flux) in this gamma-ray excess. We therefore conclude that the
gamma-ray emission more likely comes from the FSRQ
CRATES~J153246+234400.

M83. The possible source contributes to the gamma-ray excess
 is MS 13326-2935, a blazar in Veron-Cetty catalogue (also named 2E 3100),
which is 0.04$^\circ$ away from the best-fit position.
 The result is consistent with ~\citet{Lenain2011}, where a gamma-ray
 excess compatible with 2E 3100 with TS=30.1 is found.

The maximum likelihood analysis results of all the 7 starburst
galaxies in this study are shown in Table~\ref{core}. We also
perform additional maximum likelihood analysis in the energy range
from 100~{\rm MeV} to 100~{\rm GeV}, such that our results can
readily be compared with upper limits reported in ACK12. All upper
limits from this study are smaller than the corresponding ones
reported in ACK12, the latter using a smaller data set.  Note
that, for the four star-forming galaxies, M51, NGC~1365, Arp~220
and M83,  we calculated upper limits of their gamma-ray flux  by
adding a point source at the best-fit position to build a new
source background model.

\section{Interpretation and implication of the gamma-ray emission from NGC~2146}

\subsection{Interpretation of the gamma-ray emission from NGC~2146}
The gamma-ray emission of starburst galaxies could, in principle,
be produced by Bremsstrahlung and inverse-Compton emission of the
primary or secondary electrons, as well as the  pionic decay
emission resulted from CR interactions.     Detailed calculations
have shown that the  pionic decay gamma-rays dominate the emission
above 100 {\rm MeV} for starburst galaxies (e.g.
~\citet{Torres2005};~\citet{Rephaeli2010}), although leptonic
emission is expected to become increasingly important at lower
energies.

The total injected CR power in the galaxies scales with the SN
rate as SNRs are the primary CR accelerators. Following
\citet{Thompson2007}, we assume that the supernova rate  is a
constant fraction, $\Gamma_{\rm SN}$, of the star-formation rate
($\dot{M_\star}$) and a fraction of $5\eta_{0.05}\%$ of its
kinetic energy $E_{\rm SN}$ is supplied to relativistic protons,
thus the total CR power is $L_{\rm CR}=\Gamma_{\rm
SN}\dot{M_\star}\eta E_{\rm SN}$. Assuming that the star-formation
rate is related to the total infrared (IR) luminosity in 8-1000
${\rm \mu m}$ by $L_{8-1000{\rm \mu m}}=\epsilon \dot{M_\star}
c^2$,  the total CR power is (see also Eq. 5 in
\citet{Thompson2007})
\begin{equation}
L_{CR}=4.6\times10^{-4}L_{8-1000{\rm \mu
m}}E_{51}\eta_{0.05}\beta_{17},
\end{equation}
where $E_{51}=E_{\rm SN}/10^{51}{\rm ergs}$ and $\beta_{17}\equiv
(\Gamma_{\rm SN}/\epsilon)/17M\odot^{-1}$, which depends weakly on
the initial mass function \citep{Thompson2007}. In the proton
calorimeter limit, i.e. all the proton energy is lost into
secondary pions, the gamma-ray luminosity is directly related to
the CR power by $L_\gamma(>0.1{\rm GeV})\simeq\frac{1}{3}L_{\rm
CR}$, where $\frac{1}{3}$ comes from the fact that only
$\frac{1}{3}$ of the proton energy lost goes into neutral pions
and we have assumed that the energy in gamma-rays below 0.1 GeV is
negligible. Then the ratio between the total gamma-ray luminosity
and the total infrared luminosity in the calorimeter limit is
\begin{equation}
\xi\equiv\frac{L_\gamma(>0.1{\rm GeV})} {L_{8-1000{\rm \mu
m}}}=1.5\times10^{-4}E_{51}\eta_{0.05}\beta_{17},
\end{equation}
as has been predicted in \citet{Thompson2007}. We compare the
ratio between the observed gamma-ray luminosity and  total
infrared luminosity of NGC~2146 with this limit ratio in Fig.3
(together with other LAT detected galaxies) and find that
NGC~2146 lies close to this calorimeter limit. Thus,  we suggest
that NGC~2146 is probably  a "proton calorimeter".

The proton calorimetry  of NGC~2146 can be understood by comparing
the collisional energy loss time of CR protons and the escape time
\citep{Loeb2006,Thompson2007}. For a gas density of $n\sim 10 {\rm
cm^{-3}}$ in the starburst region of NGC~2146~\citep{Greve2006},
the collisional energy loss time of CR protons is
\begin{equation}
t_{pp} = (0.5n\sigma_{pp}c)^{-1}=5\times10^6 {\rm yr}
\left(\frac{n}{10{\rm cm^{-3}}}\right)^{-1},
\end{equation}
where the factor 0.5 is inelasticity, and $\sigma_{pp}$ is the
inelastic nuclear collision cross section. If $t_{pp}$ is less
than the escape time, then system is a proton calorimeter. CR
protons can escape by advection in galactic winds{\footnote{CRs
can also escape via diffusion. However, since the diffusion
coefficient in starburst galaxies is not known, this timescale is
uncertain.}}, with a timescale
\begin{equation}
t_{ad}=h/v=4\times10^6 {\rm yr} (\frac{h}{1kpc})(\frac{v}{250{\rm
km s^{-1}}})^{-1},
\end{equation}
where  $h\simeq 1{\rm kpc}$ is the size of the starburst region
~\citep{Greve2006} and $v=250-300{\rm Km s^{-1}}$ is the velocity
of the galactic wind in NGC~2146 ~\citep{Kreckel2014}. Comparing
Eqs.(3)-(4), we expect that CR protons in the starburst region of
NGC~2146 should lose a large fraction of their energy before
escape, in agreement with the above inferred large calorimetry
fraction.

\subsection{Correlation between the $\gamma$-ray luminosity and total IR luminosity }
It has been  proposed that there  is a connection between galaxy
star-formation rates and gamma-ray luminosities, motivated by the
connections between the star-formation rate, SN rate and cosmic
ray power (e.g. ~\citet{Pavlidou2002}; ~\citet{Torres2004};
~\citet{Thompson2007}; ~\citet{Stecker2007} ; ~\citet{Persic2010};
~\citet{Lacki2011}). \citet{Thompson2007} first  predicted a far
infrared (FIR)--gamma-ray correlation   in the calorimetric limit
for densest starbursts. ACK12 found a quasi-linear scaling
relation between gamma-ray luminosity and the total IR luminosity
for quiescent galaxies in the Local group and nearby starburst
galaxies. We now examine whether NGC~2146 follows this relation.
Different from ACK12, we exclude two starbursting Seyfert 2
galaxies NGC 1068 and NGC 4945 in the fit in order to have a
sample consist purely of star-forming and starburst galaxies. Fig.
\ref{correlation} shows their gamma-ray luminosities and total IR
luminosities(8-1000$\mu$m) of this sample. The total
IR(8-1000$\mu$m) luminosities are provided by~\citet{Gao2004} and
the gamma-ray luminosities are from~\citet{Ackermann2012}.  We fit
the data with a simple power-law and  find the relation
\begin{equation}
\frac{L_{0.1-100 {\rm GeV}}}{{\rm erg \ s^{-1}}}
=10^{39.772\pm0.085} (\frac{L_{8-1000\mu m}}{10^{44}\rm erg \
s^{-1}})^{1.285\pm0.058},
\end{equation}
with a Pearson correlation coefficient being $r=0.983$ and null
hypothesis probability less than $10^{-4}$, which indicates a
tight positive correlation. The scaling indices between the
gamma-ray luminosity and total IR luminosity for this sample is
consistent within uncertainties with that obtained in previous
works~\citep{Abdo2010a,Ackermann2012}.  With NGC~2146, the highest
IR luminosity galaxy in this sample, the correlation now extends
up to a total IR luminosity of $10^{11}L_\odot$, strengthening the
connection between massive star formation and gamma-ray emission
of star-forming galaxies.

\subsection{Prediction for  very high emission from NGC~2146}
Very high energy (VHE) gamma-ray emission has been detected from
nearby starburst galaxies M82 and NGC~253 by VERITAS and HESS,
respectively \citep{Acciari2009,Acero2009}. Using the GeV flux and
the spectrum of NGC~2146, we can estimate the VHE flux assuming a
simple power-law extrapolation. As can be seen from Fig.2 (left
panel), the predicted energy flux is $E^2 dN/dE\simeq10^{-13}{\rm
TeV cm^{-2} s^{-1} }(E/1{\rm TeV})^{-0.1}$ at TeV energies, which
is close to the detection sensitivity of VERITAS and should be
detectable by future instruments such as CTA, LHAASO and etc. Note
that at a distance of 15.2 Mpc, the absorption by extragalactic
background light  is unimportant for TeV photons.

Proton-proton collisions in starbursts not only produce neutral
pions, but also produce charged pions, which then decay and
produce neutrinos. \citet{Loeb2006} argued that supernova remnants
in starburst galaxies accelerate CR protons and produce
high-energy neutrinos. \citet{Loeb2006}, \citet{Thompson2007} and
\citet{Lacki2011} further made an explicit connection between the
expected diffuse gamma-ray background and the diffuse high-energy
neutrino background from starbursts since pion production links
the two directly. Recently, the IceCube Collaboration reported 37
events ranging from 60\thinspace TeV to 3\thinspace PeV within
three years of operation, correspond to a $5.7\sigma $ excess over
the background atmospheric neutrinos and
muons\citep{2014arXiv1405.5303A}. One attractive scenario for this
excess  is that they are produced by cosmic rays in starburst
galaxies\citep{2013PhRvD..88l1301M,2014PhRvD..89h3004L}. But
whether neutrinos in starburst galaxies can extend to sub-PeV/PeV
energies is uncertain, given that normal supernova remnants are
usually believed to   accelerate protons only to $\la 1{\rm PeV}$.
It was suggested that hypernova remnants in starburst galaxies, by
virtue of their fast ejecta, are able to accelerate protons to
$\gg$ PeV \citep{2007PhRvD..76h3009W} and produce sub-PeV/PeV
neutrinos \citep{2014PhRvD..89h3004L}. Future observations of VHE
gamma-ray emission of starburst galaxies extending to $\ga100$ TeV
would be able to pin down the starburst origin of the sub-PeV/PeV
neutrinos.

\section{Summary}
By searching for high-energy gamma-ray emission from the 7 best
candidate galaxies for LAT detection beyond the Local group
(ACK12), we found significant high-energy gamma-ray emission above
100 {\rm MeV} from a source spatially coincident with NGC~2146,
the nearest LIRG galaxy. The significance of the detection is
about 5.5 $\sigma$. This is the first time that a LIRG galaxy has
been detected by Fermi/LAT. The high gamma-ray luminosity suggests
that cosmic ray protons  in NGC~2146 convert most of their energy
into secondary pions, so NGC~2146 is  a  proton calorimeter. It is
also found that NGC~2146 extends the quasi-linear scaling relation
between the gamma-ray luminosities and total IR luminosities of
star-forming galaxies to a higher luminosity, thus strengthening
the connection between massive star formation and gamma-ray
emission of star-forming galaxies.

\acknowledgments We thank J. Q. Sun, J. F. Wang, K. Bechtol, C. D.
Dermer for helpful discussion. This work has made use of data and
software provided by the Fermi Science Support Center. This work
also made use of NASA's Astrophysics Data System, and the
NASA/IPAC Extragalactic Datebase(NED) which is operated by the Jet
Propulsion Laboratory, California Institute of Technology, under
contract with NASA. This work is supported by the 973 program
under grant 2014CB845800, the NSFC under grants 11273016 and
11033002, and the Excellent Youth Foundation of Jiangsu Province
(BK2012011). PHT is supported by the Ministry of Science and
Technology of the Republic of China (Taiwan) through grant
101-2112-M-007-022-MY3.

\clearpage

 \begin{figure}
    \epsscale{0.8}
  \plotone{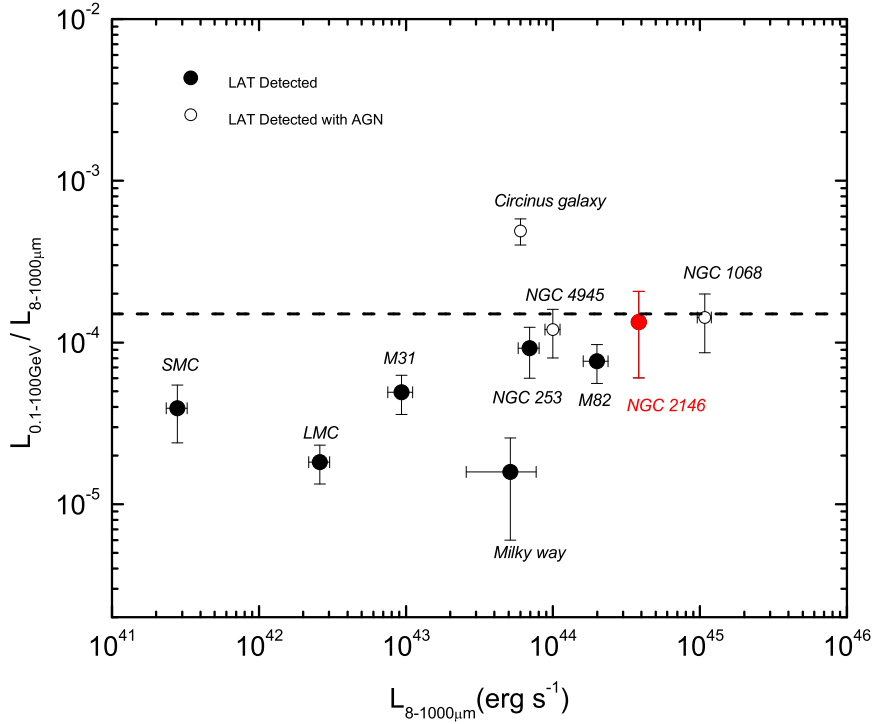}
   \caption{ The ratio between the gamma-ray luminosities and total IR luminosities for LAT-detected star-forming galaxies and
   Seyferts. The  dotted  horizontal line  corresponds to the proton calorimetric limit  (see the text for details).  The data for the Circinus galaxy, which is an  anomalously gamma-ray bright Seyfert galaxy,
   is taken from Hayashida et al. (2013).
   For other galaxies, the values of the total IR(8-1000$\mu$m) luminosities are taken from ~\citet{Gao2004} and the gamma-ray luminosities (except for NGC~2146) are taken from ~\citet{Ackermann2012}}
    \label{limit}
   \end{figure}

       \begin{figure}
    \epsscale{0.8}
  \plotone{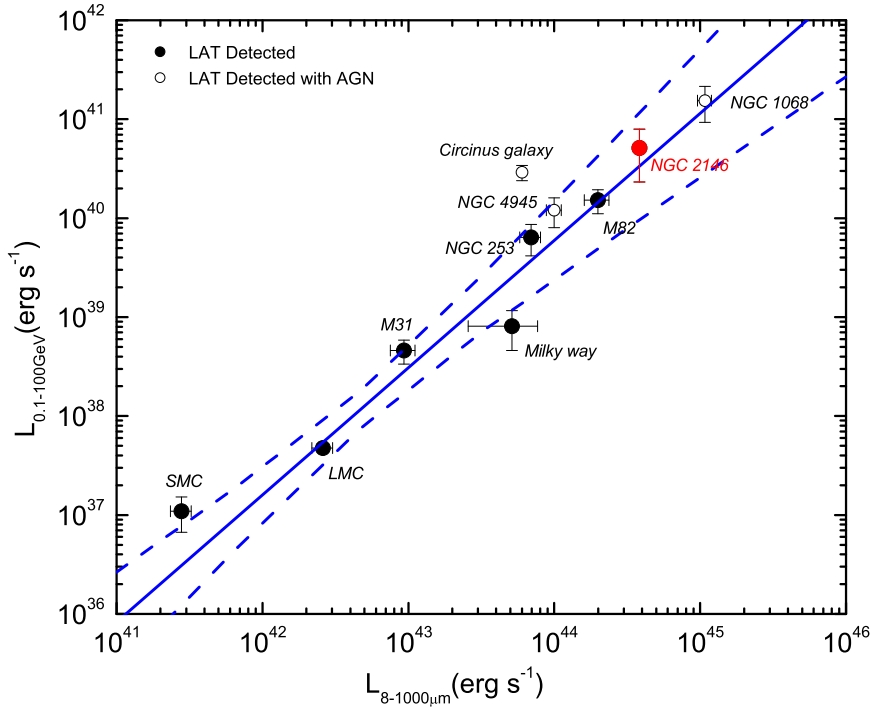}
   \caption{The relation between the gamma-ray luminosity (0.1-100 {\rm GeV}) and total IR luminosity (8-1000 $\mu$m) for star-forming galaxies.
   The red solid line is the best fit, while the blue and green dashed lines represent the upper and lower limits, respectively, at 95\% confidence.
   Note that two starbursting Seyfert galaxies (NGC 1068 and NGC 4945) and the Circinus galaxy are excluded in the fit in order to have a
sample consist purely of star-forming and starburst galaxies.}
    \label{correlation}
   \end{figure}

\clearpage
\begin{table}
\begin{center}
\caption{Best-fit position and analysis results of the five new significantly detected gamma-ray sources in the vicinity of the galaxies.\label{position}}
\begin{tabular}{crrrrrrrrrrr}
\tableline\tableline
 Galaxy        &\multicolumn{1}{c}{R.A. \& Decl.\tablenotemark{a}} & \multicolumn{1}{c}{$r_{95}$\tablenotemark{b}}    &  \multicolumn{1}{c}{ $\delta$Angle\tablenotemark{c}}  &$F(0.2-100 {\rm GeV})$   &$\Gamma$ &TS &Association\tablenotemark{d}\\
            & (deg)           & (deg)            &(deg)      & ($10^{-9}$~ph cm$^{-2}$s$^{-1}$)   &        &   \\\tableline
 NGC~2146 &(94.55, 78.30)     &0.11  &0.06  &1.1$\pm$0.5       &2.1$\pm$0.2      & 30.8&(1) \\
 M51           &(201.66, 47.81)     &0.21 &0.80  &2.8$\pm$0.8       &2.8$\pm$0.2      & 29.9 &(2)\\
 NGC~1365 &(54.17, -36.33)     &0.30   &0.65 &3.2$\pm$0.7       &2.6$\pm$0.2      & 44.4 &(3)\\
 Arp~220     &(233.24, 23.58)    &0.20   &0.46 &3.2$\pm$0.7       &2.3$\pm$0.1      & 52.2 &(4)\\
 M83           &(203.84, -29.82)    &0.07   &0.36 &1.2$\pm$0.5       &1.8$\pm$0.2      & 58.1 &(5)\\\tableline

\end{tabular}
\tablenotetext{a}{Best-fit positon of gamma-ray excess}
\tablenotetext{b}{The 95\% containment error circle radius around the best-fit position}
\tablenotetext{c}{The angular separation between the best-fit position and the position of the core of the galaxy}
\tablenotetext{d}{The most likely associated source, as shown in Sects. 2.2 and 2.3: (1) NGC~2146; (2) SDSS~J13261$+$4754; (3) PKS~0335$-$364; (4) CRATES~J153246+234400; (5) MS 13326-2935(2E 3100).}
\end{center}
\end{table}

\begin{table}
\begin{center}
\caption{Maximum Likelihood Analysis Results at the core of the starburst galaxies.\label{core}}
\begin{tabular}{crrrrrrrrrrr}
\tableline\tableline
Galaxy    &D\tablenotemark{a}      &$F(0.2-–100 {\rm GeV})$       &$\Gamma$       & $L_{0.2-100 {\rm GeV}}$    &TS &$F(0.1-100 {\rm GeV})$ \\
            &(Mpc)   & ($10^{-9}$~ph cm$^{-2}$s$^{-1}$)   &                       & ($10^{40}$~erg s$^{-1}$)  &    & ($10^{-9}$~ph cm$^{-2}$s$^{-1}$) \\\tableline
M83\tablenotemark{b}              & 3.7       &$<$1.2                 &2.2           &$<$0.3                   & ...                       &$<$2.3\\
M101                                           & 6.4       &$<$0.6                 &2.2            &$<$0.4                  & ...                       &$<$1.1\\
M51\tablenotemark{b}           &9.6          &$<$1.2                 &2.2            &$<$1.8                   & ...                       &$<$2.9\\
NGC~2146                                & 15.2     &1.1$\pm$0.6      &2.1$\pm$0.2  &4.0$\pm$2.1       & 26.4                &1.4$\pm$1.0\\
NGC 1365\tablenotemark{b}  &  20.8     &$<$0.8                &2.2             & $<$5.7                   & ...                     &$<$1.7 \\
NGC 3690                                  &  41.1    &$<$0.9                 &2.2             &$<$23.5                   & ...                     &$<$1.6\\
Arp 220\tablenotemark{b}      &  74.7     &$<$1.5                  &2.2             &$<$137.1               & ...                     &$<$3.5\\\tableline
\end{tabular}
\tablenotetext{a}{Distances are taken from~\citet{Ackermann2012}.}
\tablenotetext{b}{An additional point source at the best-fit
position was added in the source model as discussed in text.}
\end{center}
\end{table}


\begin{thebibliography}{}

\bibitem[Aartsen et al.(2014)]{2014arXiv1405.5303A} Aartsen, M.~G., Ackermann, M., Adams, J., et al.\ 2014, arXiv:1405.5303

\bibitem[Abdo et al.(2010a)]{Abdo2010a} Abdo, A.~A., Ackermann,
M., Ajello, M., et al.\ 2010a, \apjl, 709, L152


\bibitem[Abdo et al.(2010b)]{Abdo2010b} Abdo, A.~A., Ackermann,
M., Ajello, M., et al.\ 2010b, \apj, 710, 1271

\bibitem[Abdo et
al.(2010c)]{Abdo2010c} Abdo, A.~A., Ackermann, M., Ajello, M., et
al.\ 2010c, \aap, 523, L2

\bibitem[Acciari et al. (2009)]{Acciari2009} Acciari, V. A., 2009,
Nature, 462, 770

\bibitem[Acero et al. (2009)]{Acero2009} Acero, F. et al. 2009,
Science, 326, 1080

\bibitem[Ackermann et al.(2012)]{Ackermann2012} Ackermann, M.,
Ajello, M., Allafort, A., et al.\ 2012, \apj, 755, 164 (ACK12)

\bibitem[Caccianiga et
al.(2000)]{Caccianiga2000} Caccianiga, A., Maccacaro, T., Wolter,
A., Della Ceca, R., \& Gioia, I.~M.\ 2000, \aaps, 144, 247

\bibitem[Domingo-Santamar{\'{\i}}a
\& Torres(2005)]{Torres2005} Domingo-Santamar{\'{\i}}a, E., \&
Torres, D.~F.\ 2005, \aap, 444, 403

\bibitem[Gao
\& Solomon(2004)]{Gao2004} Gao, Y., \& Solomon, P.~M.\ 2004, \apj,
606, 271

\bibitem[Greve et
al.(2006)]{Greve2006} Greve, A., Neininger, N., Sievers, A., \&
Tarchi, A.\ 2006, \aap, 459, 441

\bibitem[Ginzburg
\& Syrovatskii(1964)]{Ginzburg1964} Ginzburg, V.~L., \&
Syrovatskii, S.~I.\ 1964, The Origin of Cosmic Rays, New York:
Macmillan, 1964,

\bibitem[Hayashida et al.(2013) ]{} Hayashida, M. et al. 2013, 779, 131

\bibitem[Healey et al.(2007)]{Healey2007} Healey, S.~E., Romani,
R.~W., Taylor, G.~B., et al.\ 2007, \apjs, 171, 61

\bibitem[Healey et al.(2008)]{Healey2008} Healey, S.~E., Romani,
R.~W., Cotter, G., et al.\ 2008, \apjs, 175, 97

\bibitem[Kreckel et al.(2014)]{Kreckel2014} Kreckel, K., Armus, L.,
Groves, B., et al.\ 2014, \apj, 790, 26


\bibitem[Lacki et al.(2010)]{Lacki2010} Lacki, B.~C., Thompson,
T.~A., \& Quataert, E.\ 2010, \apj, 717, 1

\bibitem[Lacki et al.(2011)]{Lacki2011} Lacki, B.~C., Thompson,
T.~A., Quataert, E., Loeb, A., \& Waxman, E.\ 2011, \apj, 734, 107


\bibitem[Lacki
\& Thompson(2013)]{Lacki2013} Lacki, B.~C., \& Thompson, T.~A.\
2013, \apj, 762, 29


\bibitem[Lenain
\& Walter(2011)]{Lenain2011} Lenain, J.-P., \& Walter, R.\ 2011,
\aap, 535, A19

\bibitem[Liu et al.(2014)]{2014PhRvD..89h3004L} Liu, R.-Y., Wang,
X.-Y., Inoue, S., Crocker, R., \& Aharonian, F.\ 2014, \prd, 89,
083004

\bibitem[Loeb \& Waxman (2006)]{Loeb2006} Loeb, A., \& Waxman, E.\
2006, JCAP, 5, 3

\bibitem[Mattox et al.(1996)]{Mattox1996} Mattox, J.~R., Bertsch,
D.~L., Chiang, J., et al.\ 1996, \apj, 461, 396

\bibitem[Murase et al.(2013)]{2013PhRvD..88l1301M} Murase, K., Ahlers, M.,
\& Lacki, B.~C.\ 2013, \prd, 88, 121301

\bibitem[Nolan et al.(2012)]{Nolan2012} Nolan, P.~L., Abdo,
A.~A., Ackermann, M., et al.\ 2012, VizieR Online Data Catalog,
219, 90031

\bibitem[Pavlidou
\& Fields(2002)]{Pavlidou2002} Pavlidou, V., \& Fields, B.~D.\
2002, \apjl, 575, L5

\bibitem[Persic
\& Rephaeli(2010)]{Persic2010} Persic, M., \& Rephaeli, Y.\ 2010,
\mnras, 403, 1569


\bibitem[Rephaeli et al.(2010)]{Rephaeli2010} Rephaeli, Y., Arieli,
Y., \& Persic, M.\ 2010, \mnras, 401, 473

\bibitem[Sanders
\& Mirabel(1996)]{Sanders1996} Sanders, D.~B., \& Mirabel, I.~F.\
1996, \araa, 34, 749


\bibitem[Stecker(2007)]{Stecker2007} Stecker, F.~W.\ 2007,
Astroparticle Physics, 26, 398

\bibitem[Thompson et al.(2007)]{Thompson2007} Thompson, T.~A.,
Quataert, E., \& Waxman, E.\ 2007, \apj, 654, 219

\bibitem[Torres(2004)]{Torres2004} Torres, D.~F.\ 2004, \apj, 617,
966

\bibitem[V{\'e}ron-Cetty
\& V{\'e}ron(2010)]{Veron2010} V{\'e}ron-Cetty, M.-P., \&
V{\'e}ron, P.\ 2010, \aap, 518, A10

\bibitem[Wang et al.(2007)]{2007PhRvD..76h3009W} Wang, X.-Y., Razzaque, S.,
M{\'e}sz{\'a}ros, P., \& Dai, Z.-G.\ 2007, \prd, 76, 083009
\end{thebibliography}
\end{document}